\documentclass{elsart5p}
\usepackage[isolatin]{inputenc}
\usepackage{graphicx}

\usepackage{amssymb}

\begin{document}

\begin{frontmatter}



\title{Preparing isomerically pure beams of short-lived nuclei at JYFLTRAP}
\author[jyfl]{T. Eronen\corauthref{cor1}}
\ead{tommi.eronen@jyu.fi},
\author[jyfl]{V.-V. Elomaa},
\author[jyfl]{U. Hager\thanksref{label2}},
\thanks[label2]{Present address: TRIUMF, 4004 Wesbrook Mall, Vancouver, British Columbia, V6T 2A3, Canada}
\author[jyfl]{J. Hakala},
\author[jyfl]{A. Jokinen},
\author[jyfl]{A. Kankainen},
\author[jyfl]{S. Rahaman},
\author[jyfl]{J. Rissanen},
\author[jyfl]{C. Weber} and
\author[jyfl]{J. Äystö}
\address[jyfl]{Department of Physics, P.O. Box 35 (YFL), FIN-40014 University of Jyväskylä, Finland}

\corauth[cor1]{Corresponding author}

\begin{abstract}
A new procedure to prepare isomerically clean samples of ions 
with a mass resolving power of more than $1\times10^5$
has been developed at the JYFLTRAP tandem Penning trap system. The method utilises a dipolar rf-excitation
of the ion motion with separated oscillatory fields
in the precision trap. During a subsequent retransfer to the
purification trap, the contaminants are rejected and
as a consequence, the remaining bunch is isomerically cleaned. This newly-developed method is suitable
for very high-resolution cleaning and is at least a factor of five faster than the methods used so
far in Penning trap mass spectrometry.

\end{abstract}

\end{frontmatter}

\section{\label{sec:intro}Introduction}
A contamination-free ion beam is desirable for any experiment with radioactive ions. 
Mass spectrometry with Penning traps, where the cyclotron frequency in a strong magnetic field
is determined \cite{bla06}, is no exception to this; contaminants do not
just deteriorate the lineshapes of the obtained spectra but also introduce systematic
frequency shifts.

Several purification steps are taken before the ions end up in the 
actual measurement stage. At the IGISOL facility in Jyväskylä \cite{ays01}, or at ISOL facilities
 in general, for instance at ISOLDE, CERN \cite{kug00}, a coarse mass
selection is performed with magnetic separators soon after extracting ions 
from the ion source. Typically a mass resolving power 
$R =\frac{M}{\Delta M_\textrm{\tiny FWHM}} = \frac{\nu}{\Delta \nu_\textrm{\tiny FWHM}} $ from~200 to~5,000
is obtained, which is adequate for a selection of a particular mass number $A$, delivering
a beam composed of isobars. These can be partially selected, for example, by chemical means or
by selective laser ionization. A mass separation in a Penning trap requires a resolving power
of $10^4$ to $10^5$. Isomeric separation is more demanding: even
 a mass resolving power of more than $10^6$ might be needed. 

The JYFLTRAP system consists of a radiofrequency quadrupole (RFQ) cooler and buncher \cite{nie02}
 and two Penning traps \cite{kol04} situated
inside the same 7~T superconducting solenoid. The RFQ is used for cooling and bunching
of the ion beam from the IGISOL separator. Isobaric as well as isomeric separation and mass measurements
are performed with the Penning traps.
The first, purification trap is used for isobaric cleaning \cite{kol04} and in most
of the cases studied, it has been sufficient to provide
contaminant-free samples for experiments. 
The second, precision trap is used for isomeric cleaning and high-precision
mass measurements.
In this article, 
complete purification schemes for separating isobars and close-by isomers
with the JYLFTRAP Penning trap setup are presented.

\section{\label{sec:trap1}Isobaric cleaning with the purification trap}
\begin{figure}[ht]
\begin{center}
\includegraphics[width=\columnwidth]{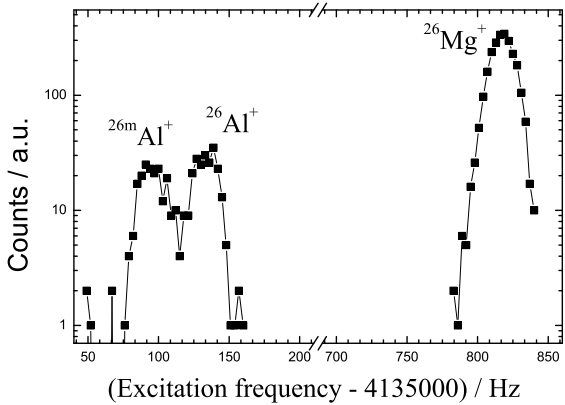}
\caption{\label{fig:mass26_trap1} Purification scan for mass $A=26$.
The excitation time that was employed was sufficient to separate the isomer and
the ground state of $^{26}$Al having different frequencies by approximately 40~Hz. }
\end{center} \end{figure}

\begin{figure}[ht]
\begin{center}
\includegraphics[width=\columnwidth]{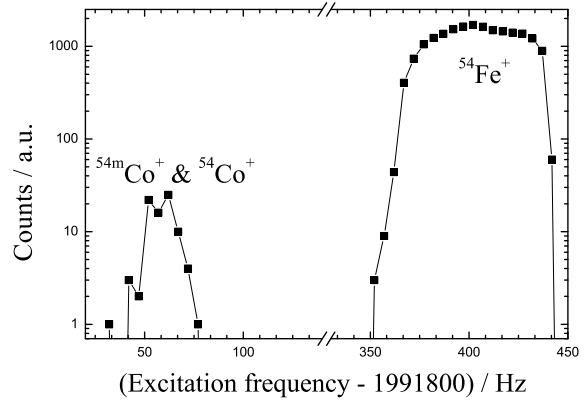}
\caption{\label{fig:mass54_trap1} Purification scan for mass $A=54$.
The isomeric and the ground state of $^{54}$Co are 
not separated. The resolution could not be improved, since the yield of stable $^{54}$Fe
is overwhelming compared with the yield of $^{54}$Co.}
\end{center} \end{figure}

The gas-filled purification trap is used to separate isobars by employing the
sideband-cooling technique \cite{sav91}. At the JYFLTRAP experiment, the following
isobaric purification scheme is used:
\begin{enumerate}
\item  An ion bunch from the RFQ is captured in the purification trap. The
injection side potential wall is lowered for a short period of time when ions are transferred into
the trap and once the ions are inside, the potential wall is restored.
\item The ions are kept in the trap for 20~to 200~ms in order to let
the axial and the cyclotron motion cool down.

\item A dipolar magnetron excitation is applied typically for 10~ms to establish a magnetron 
radius of more than 1~mm for all ions. This assures that
none of the ions can be transferred through the 2-mm orifice between the traps.

\item A mass-selective quadrupolar excitation with a frequency of $\nu_c= \frac{1}{2\pi} \frac{q}{m} B$
is applied.
This sideband excitation at the sum frequency of both radial motions $\nu_c = \nu_+ +\nu_- $
will convert the magnetron motion of the ions into a cyclotron motion, which 
is much faster cooled under the presence of a buffer gas. As a consequence, the ions
are mass-selectively driven back to the axis of the trap. 
\end{enumerate}
The achieved mass resolving power mostly depends on the buffer gas pressure and
on both the duration and the amplitude of the excitation. As long as the trap is not loaded with too many
ions, $R \le 10^5$ can be obtained.
 An example of a quadrupolar rf-frequency scan around $A=26$ in the purification
trap is shown in fig. \ref{fig:mass26_trap1}. Such a
resolving power is even enough to separate the long-lived isomer $^{26m}$Al from its ground state.
When the purification trap is loaded with more ions, the trap loses its capability to
efficiently bring the ions of interest
 back towards the axis of the trap due to increased space charge. Increasing the quadrupole amplitude and
 the buffer gas pressure enhances the centering process --- unfortunately with expense of resolution.
If the resolution is compromised (up to $\Delta \nu_\textrm{\scriptsize FWHM}$ of 100~to 200~Hz), the transmission of the
trap can be maintained. An example of a purification scan with reduced
resolution is shown in
fig. \ref{fig:mass54_trap1}. The isomer and the ground state of $^{54}$Co are not resolved,
since the difference of their cyclotron frequencies $\nu_c$ is only about 7~Hz and the trap has been tuned for better transmission
at the cost of high resolution.

\section{Isomeric cleaning with the precision trap}
Since the precision trap is situated in ultra-high vacuum ($p \le 10^{-7}~\mbox{mbar}$)
and both the cooling and
re-centering effect of the buffer gas are missing, the ion motions have to be excited differently
in order to produce clean ion samples. Here, a dipolar excitation at
the reduced cyclotron frequency $\nu_+$ is typically used.

To remove contaminant ions prior to the measurement process, their
radial amplitudes need to be increased such that they will not affect the
ions of interest. The time profile of the excitation results in a finite
lineshape in the frequency domain, given by the Fourier-transformation, where
$\Delta \nu_\textrm{\scriptsize FWHM}$ of the resonance is determined by the time
duration~$T$ of the excitation, $\Delta \nu_\textrm{\scriptsize FWHM} \propto \frac{1}{T}$.
         In addition to the normal dipolar cleaning, a Stored Waveform Inverse
Fourier Transform (SWIFT) \cite{gua96} can be used for a selective broadband cleaning.

\subsection{Dipolar cleaning with a rectangular pulse}
When using a rectangular excitation pulse, the corresponding
Fourier-transformation results in a sinc function, $\sin(x)/x$, showing
a characteristic sideband structure.
This is not inconvenient
if the frequency difference of the ion of interest and the impurity is well
known. Excitation times and frequencies need to be properly set that
contaminants are excited while the ion of interest is not. An excitation 
with a rectangular pulse is shown in fig. \ref{fig:waves} (a).
One of the frequencies for an ion of interest remaining unexcited
is marked with (2).
Despite all care to avoid excitation
of the ion of interest, there might still be some excitation, for example due to the residual
gas damping effect. On the other hand, the cleaning process with a square wave excitation is
more than factor of two faster than that with a Gaussian envelope.

\subsection{Dipolar cleaning with a Gaussian envelope} 
By using a Gaussian envelope in the excitation time profile,
the excitation pattern in the
frequency domain is also Gaussian. This way, the ion of interest just 
needs to be sufficiently separated from contaminants in order to avoid an
unwanted increase of its motional amplitudes.
 The resolution can be adjusted with the excitation time. This corresponds to 
the excitation pattern (b) shown in fig. \ref{fig:waves}. The contaminant, 
marked (0), is excited maximally while the ion of interest (3) remains
unaffected. 

This method is routinely used in many trap setups like ISOLTRAP \cite{bol96} at CERN, CPT \cite{kla03} at
ANL and LEBIT \cite{sch05} at MSU. The advantage in using dipolar cleaning is that it
can be applied either in low- or high-resolution modes: Using a
very short duration and a high amplitude will remove ions
within a bandwidth of several kHz. Or, the amplitude can be set low and the duration
long to obtain a narrow-band cleaning.

ISOLTRAP has demonstrated a selection of nuclear isomers
\cite{bla05b} in combination with a selective laser ionisation and decay spectroscopy. Here,
states of $^{70}$Cu were separated 
with a Penning trap.
In the cleaning process, a mass resolving power of about~$2\times 10^5$ was obtained
with a 3-s excitation time. Although there was only one contaminant to be cleaned
with the Penning trap,
it should be noted that usually several contaminants are cleaned in parallel.

\subsection{\label{sec:rclean}Dipolar cleaning with separated oscillatory fields}
\begin{figure}[ht]
\begin{center}
\includegraphics[width=\columnwidth]{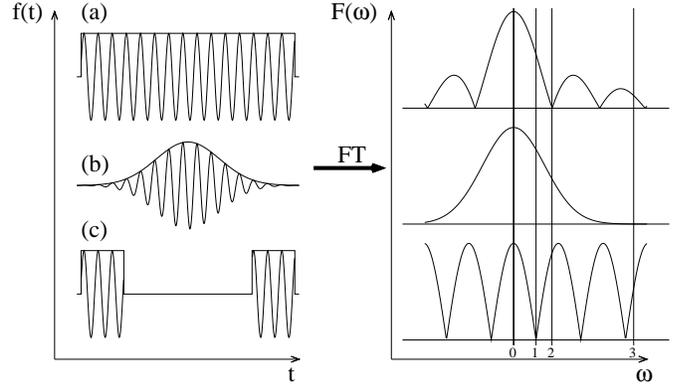}
\caption{\label{fig:waves}
Examples of excitation time profiles $f(t)$ (left side) for (a) a rectangular
excitation pulse, (b) a Gauss-modulated envelope and (c) an excitation with time-separated oscillatory fields. The corresponding Fourier transformations
$F(\omega)$ in the frequency domain are shown on the right. The position (0)
indicates the frequency of the contaminant ion species and (1) to (3) indicate a possible 
frequency of an ion of interest. For the discussion see text.
}
\end{center} \end{figure}

The resolution can be further enhanced by using an excitation scheme
with time-separated oscillatory fields \cite{ram90,bol92,kre07}. It has been shown
that the linewidth is reduced by almost 40~\% \cite{geo07}.
The reduction
is illustrated in fig. \ref{fig:waves}, comparing the width of the main peak for (a) and (c). 
In case (c) the ion of interest remains
unexcited at the frequency position (1).
As in the case of a rectangular excitation profile, 
both frequencies have to be known accurately beforehand and can
be conveniently used for example in measurements of superallowed beta-emitters or stable ions.
This method is less suited for studying short-lived nuclides
with unknown mass values.

\section{Isomeric cleaning with time-separated oscillatory fields and additional cooling}
\begin{figure}[ht]
\begin{center}
\includegraphics[width=\columnwidth]{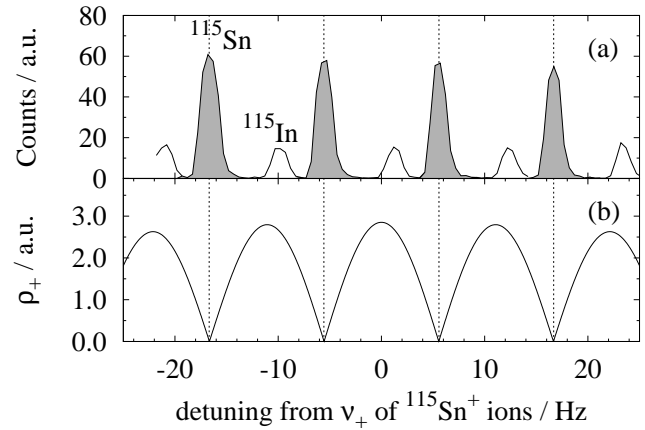}
\caption{\label{fig:ramsey_cleaning_scan} 
Number of detected ions at the microchannel-plate detector behind the trap setup as a function of
the applied dipolar frequency in the precision trap (a). Here, an excitation time
pattern of (10-80-10)~ms (On-Off-On) was used. Prior to their ejection the
ions have been extracted backwards, captured and re-centered in the purification trap.
The individual peaks are the transmitted ions of $^{115}$Sn (grey) and $^{115}$In 
(white). The lower panel (b)
shows the expected increase of the cyclotron radius $\rho_+$ for $^{115}$Sn as a
 function of the applied dipolar
frequency. Here, the highest transmission occurs when least excited.}
\end{center} \end{figure}

In the methods introduced so far, the cleaning is performed in such a way
that the contaminants are driven to orbits with large radii such that they
might hit the ring electrode. 
To avoid excessive excitation, the ion sample can be extracted towards the purification trap: The 
contaminants will hit the electrode surface
surrounding the 2-mm diaphragm between the traps, while the ions of interest can pass through.
The remaining cleaned bunch of ions is then captured in the purification trap, where
 the ions are additionally recooled and recentered as described in section \ref{sec:trap1}. 
In this way, it is possible
to perform a very-high resolution cleaning in a time-efficient manner.
For instance, the mass doublet $^{115}$Sn and $^{115}$In has a
cyclotron frequency difference of about 4.5~Hz. Using an excitation
time pattern of (10-80-10)~ms (On-Off-On) the different 
isotopes are fully separated. An example of a transmission curve after the second
recentering in the purification trap
is shown in fig. \ref{fig:ramsey_cleaning_scan}.

\begin{figure}[ht]
\begin{center}
\includegraphics[width=\columnwidth]{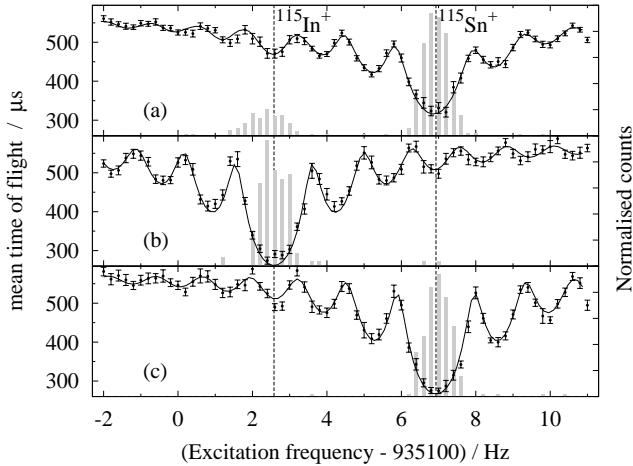}
\caption{\label{fig:both_and_cleaned} Time-of-flight ion cyclotron
resonances and a two-peak fit for an uncleaned ion sample (a). 
The $^{115}$In resonance is barely visible. In the middle panel (b),
$^{115}$Sn has been cleaned away, which remarkably enhances the $^{115}$In resonance.
In the bottom panel (c), $^{115}$In has been removed from the precision trap.
The vertical bars indicate the normalised number of ions having a time of flight less than 350~$\mu$s.
}
\end{center} \end{figure}

After a repeated centering and cooling in the purification trap the ion
bunch is transferred to the precision trap for the actual cyclotron frequency
determination. Figure \ref{fig:both_and_cleaned} demonstrates the prospects of this cleaning
procedure applied to the $A = 115$ isobars of indium and tin. If no cleaning is
applied, $^{115}$In is barely visible in the cyclotron resonance curve (a).
The ratio between the detected ions of $^{115}\mbox{Sn}$ to
$^{115}\mbox{In}$ is 5 to 1. The number can be quantified when selecting
only those ions with a time of flight shorter than 350~$\mu$s, which were
affected by a resonant excitation at their respective cyclotron
frequency $\nu_c$. The corresponding number of detected ions is indicated by
vertical bars.
In (b) $^{115}$Sn was removed by a dipolar excitation at $\nu_+$ in the
precision trap, as described in section \ref{sec:rclean}, resulting
in clean resonance of $^{115}$In. In (c), $^{115}$In was removed in analogy
to case (b).

\section{Application to superallowed $Q_\textrm{\scriptsize EC}$-value measurements of $^{50}$Mn and $^{54}$Co}
The motivation to develop a fast and universal cleaning scheme arises from a recent 
proposal for $Q_\textrm{\scriptsize EC}$ measurements for $^{50}$Mn
and $^{54}$Co. Both nuclides have relatively short half-lives
$T_{1/2}$ of 300~ms and 200~ms, respectively. In addition,
both have long-lived isomers with low excitation energies of
about $200~\mbox{keV}$.
As can be seen from fig. \ref{fig:mass54_trap1}, the states are not resolved
with the conventional experimental sequence
in the purification trap. 

\begin{figure}[ht]
\begin{center}
\includegraphics[width=\columnwidth]{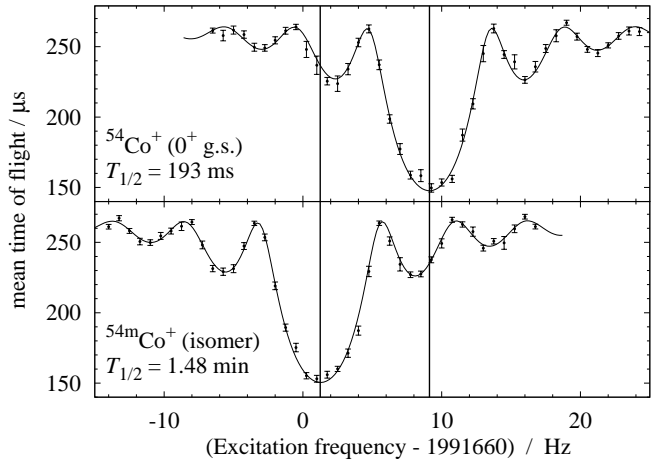}
\caption{\label{fig:rclean_co}Time-of-flight ion cyclotron resonances for ground and isomeric state of $^{54}$Co
measured in November 2006.
The unwanted states have been cleaned using dipolar excitation with separated oscillatory 
fields. For the cyclotron frequency determination an excitation time of 200~ms was used.
}
\end{center} \end{figure}

The fastest way of separating the ground and the isomeric state is to use dipolar cleaning with
separated oscillatory fields.
For the dipolar excitation, a time of 80 ms and for the repeated cooling and
centering of the ions in the purification trap 100 ms are required. The
resulting decay losses are about 50~\%, which are counterbalanced by a rather
high production rate.
Examples of time-of-flight resonances are shown in fig. \ref{fig:rclean_co}.

\begin{figure}[ht]
\begin{center}
\includegraphics[width=\columnwidth]{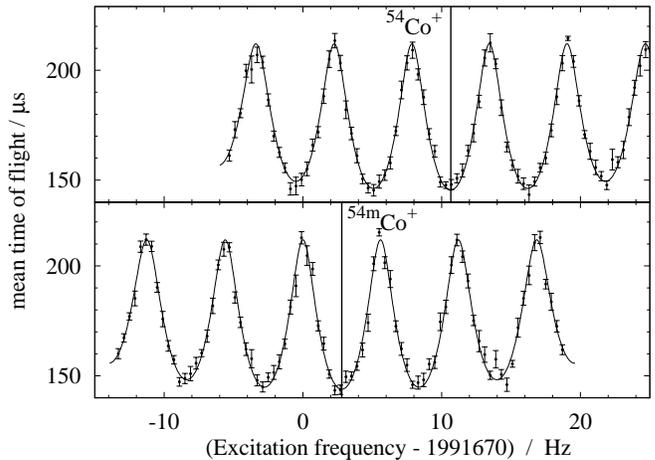}
\caption{\label{fig:rclean_rtof}Time-of-flight ion cyclotron resonances for ground and
isomeric state of $^{54}$Co
using separated oscillatory fields for both the dipolar cleaning as well as 
for the cyclotron resonance measurement. 
A dipolar excitation pattern of (10-55-10)~ms and a quadrupolar
excitation pattern of (25-150-25)~ms were used.
The data were obtained in May 2007.
}
\end{center} \end{figure}

The excitation with separated oscillatory fields has
been demonstrated to be well suited also for time-of-flight
cyclotron resonances \cite{geo07}. The theoretical lineshape for fitting purposes is extensively
described in \cite{kre07}.
Figure \ref{fig:rclean_rtof} shows an example
where both the cleaning and the time-of-flight resonance were produced with an excitation by separated
oscillatory fields. 

The linewidths of the cyclotron resonances shown in figs. \ref{fig:rclean_rtof} and
\ref{fig:rclean_co}  are
5.9 Hz and 3.7 Hz, respectively. In both cases a total time duration of
200~ms was spent in the excitation process. The resonances 
obtained with time-separated oscillatory fields are 35~\% narrower.
The uncertainty in the determination of the line center is reduced by a factor of
2.5.

\section{Conclusions}
A new high-resolution cleaning scheme employing excitation with time-separated oscillatory
fields has been demonstrated. Unwanted isobaric or isomeric contaminants will be
completely removed after the cleaning excitation when the ions are retransferred from the precision
to the purification trap through the narrow channel. Once
the cleaned ion sample has been recooled and recentered in the purification trap, the sample
is ready for extraction to the experiments. Recently, this newly-developed cleaning
method has been succesfully used to prepare clean samples of the ground states of
$^{50}$Mn and $^{54}$Co, where a mass resolving power of $R=3\times 10^5$ was needed.
This cleaning method is now routinely used at JYFLTRAP to provide isomerically
clean beams for experiments.

\section*{Acknowledgements}

This work has been supported by the EU's 6th
Framework Programme ``Integrating Infrastructure
Initiative -- Transnational Access'' Contract Number:
506065 (EURONS, Joint Research Activities
TRAPSPEC and DLEP) and by the Academy of
Finland under the Finnish Centre of Excellence Programmes
2000-2005 (ProjectNo. 44875, Nuclear and
Condensed Matter Physics Programme) and 2006-2011
(Nuclear and Accelerator Based Physics Programme at JYFL).


\end{document}